\documentstyle[12pt]{article}
\def\comma      { \, , }
\def\period     { \, . }
\def\bra#1      { \langle \, #1 \, \vert \, }
\def\ket#1      { \, \vert \, #1 \, \rangle \, }
\def\semiket#1  { \,  \, #1 \, \rangle \, }
\def\del        {  \partial  }
\def\half       {  {1\over 2}  }
\def\abs#1      {  \, \vert #1 \vert \,   }
\def\eqabegin         {  \begin{eqnarray}  }
\def\eqaend           {  \end{eqnarray}  }
\def\nn               {  \nonumber  }
\def\parn              {  \par\noindent }
\def\parbigskip        {  \par\bigskip  }

\def\titleandfile#1#2
{  \begin{center}{\Large\bf #1}\end{center}
              \par\begin{flushright} #2 \end{flushright}  }
\def\sectionnumbering { \setcounter{equation}{0}
   \renewcommand{\theequation}{\arabic{section}.\arabic{equation}}}
\def\appendixnumbering#1 { \setcounter{equation}{0}
         \renewcommand{\theequation}{\arabic{#1}.\arabic{equation}}}
\def\msection#1      {  \begin{center} \section{#1} \end{center}   }
\def\nsection#1      {  \let\boldface\bf \def\bf{} \section{#1}
                           \let\bf\boldface   }
\def\mnsection#1     {  \begin{center} \nsection{#1} \end{center}  }
\def\csection#1
  { \begin{center} \let\boldface\bf \def\bf{\large\sc}
     \section{#1} \let\bf\boldface \end{center} \sectionnumbering }
\def\csectionast#1
  { \begin{center} \let\boldface\bf \def\bf{\large\sc}
     \section*{#1} \let\bf\boldface \end{center} \sectionnumbering }
\def\csubsection#1
      {  \let\boldface\bf \def\bf{\normalsize\sc} \subsection{#1}
                           \let\bf\boldface   }
\newcommand{\nullify}[1]{}
\def\short       {$\!\!$}
\def\NPB#1       {Nucl. Phys. {\bf B#1} \short }
\def\PLB#1        {Phys. Lett. {\bf B#1} \short }
\def\PRL#1       {Phys. Rev. Lett. {\bf #1} \short }
\def\PR#1        {Phys. Rev. {\bf #1} \short }
\def\PRD#1       {Phys. Rev. D {\bf #1} \short }
\def\AnnPhys#1   {Ann. Phys. {\bf #1} \short }
\def\CMP#1       {Commun. Math. Phys. {\bf #1} \short }
\def\JMP#1       {J. Math. Phys. {\bf #1} \short }
\def\JPA#1       {J. Phys. A {\bf #1} \short }
\def\IJMP#1      {Int. J. Mod. Phys. {\bf #1} \short }
\def\MPL#1       {Mod. Phys. Lett. {\bf #1} \short }
\def\JETP#1      {Sov. Phys. JETP {\bf #1} \short }
\def\coset {SL(2,{\bf R})/U(1)}
\def\sltwo   {SL(2,{\bf R})}
\def\lp    {l_p}
\def\Epsilon   {{\cal E}}
\def\calhD {\hat{{\cal D}}}
\def\hatxi {\hat{\xi}}
\def\hath  {\hat{h}}
\def\hatA  {\hat{A}}
\def\hatg  {\hat{g}}
\def\hatD  {\hat{D}}
\def\hatR  {\hat{R}}

\begin{document}
%%%%%%%%%%%%%%%%%%%%%%%%%%%%%%%%%%%%%%%%%%%%%%%%%%%%%%%
%%%%%%%% cover %%%%%%%%%%%%%
%%%%%%%%%%%%%%%%%%%%%%%%%%%%%%%%%%%
\def\papertitlepage{\baselineskip 3.5ex \thispagestyle{empty}}
\def\Title#1{\baselineskip 1cm \vspace{1.5cm}\begin{center}
 {\Large\bf #1} \end{center}
\vspace{0.5cm}}
\def\Authors#1{\begin{center} {\rm #1} \end{center}}
\def\Abstract{\vspace{1.0cm}\begin{center} {\large\bf Abstract}
           \end{center} \parbigskip}
\def\Komabanumber#1#2#3{\hfill \begin{minipage}{4.2cm} UT-Komaba #1
              \parn #2
              \parn #3 \end{minipage}}
\renewcommand{\thefootnote}{\fnsymbol{footnote}}
%%%%%%%%%%%%%%%%%%%%%%%%%%%%%%%%%%%%%%%%%%%%%%%
\renewenvironment{thebibliography}{\pagebreak[3]\par\vspace{0.6em}
\begin{flushleft}{\large \bf References}\end{flushleft}
\vspace{-1.0em}
\renewcommand{\labelenumi}{[\arabic{enumi}]\ }
\begin{enumerate}\if@twocolumn\baselineskip=0.6em\itemsep -0.2em
\else\itemsep -0.2em\fi\labelsep 0.1em}{\end{enumerate}}
%%%%%%%%%%%%%%%%%%%%%%%%%%%%
\baselineskip=0.7cm
%%%%%%%%%%%%%%%%%%%
\papertitlepage
\vspace*{0cm}
\Komabanumber{95-11}{hep-th/9512019}{December 1995}
%%%%%%%%%%%%%%%%%%%%%%%%%%%%
\Title{Field Theoretical Quantum Effects \\ on \\ the Kerr Geometry}
\Authors{{\sc
Y.~Satoh
\footnote[2]{e-mail address:\quad
ysatoh@hep1.c.u-tokyo.ac.jp}
 \\
}
\vskip 3ex
 Institute of Physics, University of Tokyo, \\
 Komaba, Meguro-ku, Tokyo 153 Japan \\
  }
\vspace{-0.5cm}
%%%%%%%%%%%%%%%%%%%%%%%%%%%%%%%%%%%%%%%
\Abstract \noindent
\baselineskip=0.7cm
%%%%%%%%%%%%%%%%%%%%%%%%%%%%%%%%%%%%%%
We study quantum aspects of the Einstein gravity
with one time-like and one space-like Killing vector commuting
with each other. The theory is
 formulated as a $\coset$ nonlinear $\sigma$-model
 coupled to gravity.
The quantum analysis of the nonlinear $\sigma$-model
part, which includes all the dynamical degrees of freedom,
can be carried out in a parallel way to ordinary nonlinear
$\sigma$-models in spite of the existence of an unusual
coupling. This means that we can investigate consistently
the quantum
properties of the Einstein gravity, though we are limited
to the fluctuations depending only on two coordinates.
We find the forms of the beta functions to all orders up to numerical
coefficients. Finally we consider the quantum effects of the
renormalization on the Kerr
black hole as an example. It turns out that the asymptotically
flat region remains intact and stable, while,
in a certain approximation,
it is shown that the inner geometry changes considerably however
small the quantum effects may be.
%%%%%%%%%%%%%%%%%%%%%%%%%%%%%%%%%%%%%%%
\parbigskip
%%%%%%%%%%%%%%%%%%%%%%%%%%%%%%%%%%%%%
PACS number(s): 04.06.+n
\newpage
\renewcommand{\thefootnote}{\arabic{footnote}}
\setcounter{footnote}{0}
%%%%%%%%%%%%%%%%%%%%%%%%%%%%%%%%%%%%%%%
%%%%%%%%%%%%%%%%%%%%%%%%%%%%
\baselineskip=0.7cm
%%%%%%%%%%%%%%%%%%%%%%%%%%%%
%%%%%% introduction %%%%%%%%
\csection{Introduction}
Of quantum aspects of Einstein gravity we are far from a clear
understanding
at present. There exist many difficulties both conceptually
and technically.
One of the most outstanding ones is its nonrenormalizability.
Due to this,
we have no consistent way to investigate its quantum theory.
In spite of the difficulties, many attempts have been made
for incorporating quantum effects, for example,
by using semiclassical and $1/N$ approximations.
In these approaches, the flat space-time is suffered from
the instability
by the quantum perturbation owing to the induced higher
derivative terms or the tachyonic modes in the gravitational
propagator \cite{Horowitz}-\cite{Simon}.
Moreover, the tachyonic modes make the actual perturbative
calculation impossible.
The theories with higher derivative terms are
studied also as an effective ones in the low energy limit
of some fundamental theory such as string theory
(see, e.g., \cite{BD}).
In these theories, the higher derivative terms appear
in the perturbation with respect to weak curvature.
Hence we cannot deal with the region with strong curvature where
quantum effects are expected to become important, and we can
consider only small
deviations from the classical solutions.
There is also ambiguity related to field redefinitions
(\cite{Natsuume}, for instance).
In a much more simplified setting, the quantum mechanics of
minisuperspace \cite{HH}
or the Schwarzschild black hole has been investigated \cite{Kuchar}.
In these cases, it is still difficult to extract their
physical consequences. Thus we have not yet succeeded in grasping
definite quantum aspects of general relativity even
in some approximation.

Difficulties concerning quantum Einstein gravity are expected to be
overcome when we understand a more fundamental theory. Intensive
studies
have been made in this direction, but, together with this, it may be
important to
accumulate certain pieces of knowledge of quantum properties of
Einstein gravity
even if in a simplified setting.

In this article, we shall work with the Einstein gravity with
one time-like
and one space-like Killing vector commuting with each other.
The Einstein gravity with two commuting Killing vectors
can be formulated as a $\coset$ nonlinear $\sigma$-model coupled
to gravity \cite{HD,Nicolai}.
For the Einstein-Maxwell system, we have a similar formulation as a
$SU(2,1)/SU(2)\times U(1)$ or $SU(2,1)/SU(1,1)\times U(1)$ nonlinear
$\sigma$-model according to the signatures of the Killing vectors.
One of the most important applications of these facts is the proof
of the
uniqueness theorem of the Kerr-Newman solution by \cite{MB}.
The central equation of these systems is known as the Ernst equation
\cite{Ernst}.
For generating the exact solutions, these
systems have been studied extensively
and many interesting and rich structures have been
revealed \cite{HD,Nicolai}, \cite{KSHM}.
In particular, the systems possess infinite dimensional hidden
symmetries \cite{Geroch}-\cite{BM} and become integrable
\cite{BZ,Maison}. In addition, the similarity between these hidden
symmetries and those
of dimensionally reduced supergravities has been
recognized \cite{Julia},
and recently applied to the study of string dualities \cite{SKM}.

As for nonlinear $\sigma$-models, there exists an extensive
literature on
its quantum analysis \cite{PZA}-\cite{DMV}.
In two dimensions, nonlinear $\sigma$-models are renormalizable
(in a generalized sense by Friedan \cite{Friedan}) and
their quantum aspects can be studied in a consistent way at least
perturbatively. Furthermore, among various models, the simplest
one is the $O(3)$ or ${\bf C}P^1(SU(2)/U(1))$ nonlinear
$\sigma$-model,
and its target manifold $ {\bf C}P^1 $ is the compact analog
of our $\coset$.

Therefore we can expect to make use of the vast knowledge
in the literatures.
The purpose of this article is to study the quantum theory of
the Einstein
gravity reduced to two dimensions and to investigate its effects
on geometry.
Of course, in our formulation in which some of the quantum
fluctuations
are truncated, we can say only a little about the statistical
aspects of the original Einstein gravity.
However it turns out that we can actually deal with the quantum
theory of this reduced Einstein gravity and evaluate some effects on
geometry in a consistent and simple way.
We believe that our analysis gives some insights into
quantum aspects of Einstein gravity.

The rest of this article is organized as follows.
In Sec.2, we formulate the Einstein gravity with the two Killing
vectors as a $\coset$ nonlinear $\sigma$-model and equations
governing the system
are derived. Next, in Sec.3, we investigate
the quantum theory of the nonlinear $\sigma$-model part.
The beta functions are obtained to all orders up to numerical
coefficients determined by explicit loop calculations. Then the
equations
including the renormalization effects are given. Sec.4 is devoted to
the analysis of the quantum effects on the Kerr black hole.
We find that the asymptotically flat region remains intact and
stable.
On the other hand, in a certain approximation at one loop order,
it is shown that the inner geometry undergoes a considerable change
no matter how small the quantum effects may be. Finally,
brief discussion is given in Sec.5. Throughout this article,
we adopt the sign convention in which the flat space-time metric
in four dimensions is $\eta_{MN} =$ diag$(-1,1,1,1)$.
%%%%%%%%%%%%%%%%%%%%%%%%%%%%
%%%%%% section 2 %%%%%%%%%%%
\csection{Dimensionally Reduced Einstein Gravity}
In this section, we consider the dimensional reduction of
the Einstein gravity with two commuting Killing vectors.
By
dropping the dependence on the direction of one isometry, and
performing a dual transformation,
we find that
the theory is described by a $ \coset $ nonlinear
$\sigma$-model coupled to gravity in three dimensions.
Then we further reduce the theory to two dimensions.
We shall follow
the method adopted in \cite{Nicolai}, and deal with the case
in which one Killing vector is time-like and the other is space-like.

We begin with the following vierbein in a triangular gauge,
\eqabegin
   E_{\ M}^A &=& \left(
                 \begin{array}{cc}
                     \Delta^{-1/2} \ e_m^a & \Delta^{1/2} A_m \\
                     0                     & \Delta^{1/2}
                 \end{array}
             \right)
   \comma
\eqaend
where $M (= 0 - 3)$ and $m (= 1 - 3)$ refer to the space-time indices
and  $A (= 0 - 3)$ and $a (= 1 - 3)$ to those of its tangent space.
Assuming that all the components are independent of the time-like
coordinate, $x^0$,  the Einstein-Hilbert action is reduced to
\eqabegin
 \frac{1}{\hbar} \ S_{EH}
     &=& \frac{1}{\hbar \kappa} \int d^4 x \ E  R^{(4)}(E) \nn \\
     &=& \frac{L}{\lp^2}
          \int d^3x \ e \left[ \ R^{(3)}(e)  + \frac{1}{4}
          \Delta^2 F_{mn}F^{mn} - \half \gamma^{mn} \Delta^{-2}
          \del_m \Delta \del_n \Delta \right]
     \comma \label{S31}
\eqaend
where $\kappa$ is given by $\kappa = G/c^3$, $\lp$ is the Planck
length, $L$ is the "length" of $x^0$ direction,
and $E$ and $e$ are det$ \ E_M^A$ and det$ \ e_m^a$, respectively.
 $F_{mn}$ is
defined by $F_{mn} = (d A)_{mn} \equiv \del_m A_n - \del_n A_m $
 and the indices are raised
and lowered by the three metric, $\gamma_{mn}$, determined
by the dreibein $e_m^a$.

The equations of motion derived from the above reduced action
have a $\sltwo$
symmetry. Although it is not manifest in
Eq.(\ref{S31}), we can obtain the action manifestly
invariant under this symmetry by a dual transformation
\footnote{The author would like to thank Y. Kazama for the
discussion on this point.}.
First, let us introduce an auxiliary field, $C_{mn}$, add a term
to the Lagrangian, ${\cal L}$, as
\eqabegin
   {\cal L} &\to &
       {\cal L} + C^{mn} \left[ F_{mn} -  (d A)_{mn} \right]
    \comma
\eqaend
and regard $F_{mn}$ as an
independent field. By integrating out $C_{mn}$, we get
$F_{mn} = (d A)_{mn}$
and the original action. On the other hand, the integration of
$A_m$ leads
to $\nabla_m C^{mn} = 0 $ and hence $C_{mn}$ can be written by a
scalar field $B$ as $C^{mn} = \half \epsilon^{mnl} \del_l B $, where
$\nabla_m$ and $\epsilon_{mnl}$ are the covariant derivative operator
and the volume element, respectively. Finally, by the further
integration of $F_{mn}$, we get the following Lagrangian,
\eqabegin
   e \ {\cal L}^{(3)} &=&
        e \  \left[\ R^{(3)} (e) - \half \gamma^{mn} \Delta^{-2}
              \left(  \del_m B \del_n B
                      + \del_m \Delta \del_n \Delta \right)
          \  \right]
    \comma
\eqaend
and the relation between $F_{mn}$ and $B$,
\eqabegin
   \Delta^2 F_{mn} &=& - \epsilon_{mnl}\  \del^l B
   \period \label{dual}
\eqaend
We can check that the model obtained in this way is actually
equivalent on-shell to the original one.
As intended, ${\cal L}^{(3)}$ has a $\sltwo$ symmetry,
\eqabegin
  Z \longrightarrow \ Z' &=& \frac{a Z + b}{c Z + d} \ ; \quad
     \left(
        \begin{array}{cc}
           a & b \\
           c & d
        \end{array}
     \right) \ \in \sltwo
   \comma
\eqaend
where $Z = B + i \Delta $.
$Z$ is related to the so-called Ernst potential $\Epsilon$ by
\eqabegin
   \Epsilon &=& i \bar{Z} \ = \ \Delta + i B
   \period
\eqaend
Moreover we find that
the model described by ${\cal L}^{(3)}$ is a
$\coset$ nonlinear $\sigma$-model (coupled to gravity), and this is
a non-compact analog of a $O(3)$ $({\bf C}P^1)$ nonlinear
$\sigma$-model. The analogy becomes obvious in the forms,
\eqabegin
  &&    - \half \gamma^{mn} \Delta^{-2}
             \left(  \del_m B \del_n B
                  + \del_m \Delta \del_n \Delta \right)
       \\
   && \qquad \quad
       = \  - \half \gamma^{mn} \Delta^{-2}
        \del_m \Epsilon \del_n \Epsilon
      \ = \
    - 2 \gamma^{mn}
         \frac{\del_m w \del_n \bar{w}}{(1- w \bar{w})^2}
   \ = \    - \half \gamma^{mn}\ \eta^{ab} \del_m v^a \del_n v^b
          \nn \comma
\eqaend
where $(w-1)/(w+1) = \Epsilon $, and $v^a$ is
defined by $w =(i v^1 - v^2)/(1 + v^0)$,
$\eta_{ab} v^a v^b = -1$, and
$\eta_{ab}=$diag$(-1,1,1)$ \cite{Gruszczak}.

Now let us reduce the theory further to two dimensions.
As in the previous case, we take the following form of the
dreibein in a triangular gauge,
\eqabegin
  e^a_{\ m} &=& \left(
               \begin{array}{cc}
                  \lambda \delta_\mu^\alpha
                    & \rho C_\mu    \\
                  0 & \rho
               \end{array}
            \right) \label{e2}
   \comma
\eqaend
where $\mu, \alpha = 1,2$. Since $C_\mu$ has no physical degrees of
freedom, we can set $C_\mu = 0$.
Then by dropping the dependence on $x^3$,  we obtain
\eqabegin
  \frac{1}{\hbar} S_{EH}  &=&
       \frac{V}{\lp^2} \int d^2 x \ e \ {\cal L}^{(2)}
    \comma  \label{S2} \\
  e \ {\cal L}^{(2)} &= &
     \ e \left[ R^{(2)}
       - \half \gamma^{\mu\nu} \Delta^{-2} \del_\mu
   \Epsilon \del_\nu \bar{\Epsilon}
                \right]
    \  = \
 \rho \ \delta^{\mu \nu} \left[ -2 \del_\mu \del_\nu \ln \lambda
     - \half  \Delta^{-2} \del_\mu \Epsilon \del_\nu \bar{\Epsilon}
                 \right]
 \comma \nn
\eqaend
where $V$ is the "volume" of $(x^0,x^3)$ space-time. Note that, in
the latter form, the indices are contracted effectively by
the flat two dimensional metric $\delta_{\mu \nu}$. Thus
in the following, it is understood that the
indices are raised and lowered by the flat metric.

The independent equations of motion deduced from the above action are
\eqabegin
 &&\del_\mu \del^\mu \rho = 0 \label{eqrho}\\
 && \Delta \del^{\mu} \left(  \rho \del_\mu  \Epsilon \right)
      =   \rho  \ \del_\mu \Epsilon \del^\mu \Epsilon
      \label{Ernsteq}\\
 && \del_\zeta \rho \del_\zeta \ln \lambda - \half \del_\zeta^2 \rho
  = \frac{1}{4} \ \rho \ \Delta^{-2}
   \del_\zeta \Epsilon \del_\zeta \bar{\Epsilon}
 \comma \label{emconst}
 \eqaend
where $\zeta = x^1 + i x^2 $.
These equations are derived by the variations of $\Epsilon$ and
$\gamma_{\mu\nu}$.
The variation of $\rho$ leads to a dependent equation.
This is related to the
fact that the degree of freedom of $\rho$ is spurious. Indeed since
$\rho$ is a free field and there remains the choice of the conformal
gauge in two dimensions preserving the form of $e^a_{\ m}$
in (\ref{e2}),
we can identify $\rho$ with one of the coordinates by some
conformal transformation.

In the reduced theory to two dimensions, (\ref{dual}) leads to
$A_{1,2} = 0 $ and
\eqabegin
  \Delta^2 \del_\zeta A &=&  i  \rho  \del_\zeta B
   \period \label{dual2}
\eqaend
Consequently, we have four basic equations, Eq.(\ref{eqrho})
$-$ Eq.(\ref{dual2}).
Eq.(\ref{Ernsteq}) for
$\Epsilon$ is known as the Ernst equation and
becomes integrable if we set $\rho$ to be one of the coordinates.
There exists vast knowledge of this equation.
For detail, see \cite{HD,Nicolai}, \cite{KSHM}.

As for the metric in four dimensions,
in our parametrization we have
\eqabegin
  ds^2 &=& \gamma_{MN} dx^M dx^N
       \ = \ \eta_{AB} E_{\ M}^A E_{\ N}^B \ dx^M dx^N \nn \\
       &=& \Delta^{-1}
         \left[ \lambda^2 \left( (dx^1)^2 + (dx^2)^2 \right)
                + \rho^2 (dx^3)^2  \right]
        - \Delta \left( dx^0 + A \ dx^3 \right)^2
    \period \label{4metric}
\eqaend

%%%%%%%%%%%%%%%%%%%%%%%%%%%%
%%%%%% section 3 %%%%%%%%%%%
\csection{Renormalization of Noninear $\sigma$-model Part}
In the previous section, we formulated the Einstein gravity with
two commuting Killing vectors as a two dimensional $\coset$ nonlinear
$\sigma$-model coupled to gravity.  In Sec.3, we consider the
quantization of the nonlinear $\sigma$-model part which includes
all the dynamical degrees of freedom.
This means that we investigate
the effects of the quantum fluctuations maintaining the symmetry
of the isometries
(independence of $x^0$ and $x^3$).
Because the three dimensional gravity part
($\lambda \comma \rho$ etc.)
has no physical degrees of freedom, we can expect that it does not
make main contributions to the quantum effects. Thus we left
the quantization of this part as a future problem.
In quantum theory, it is ambiguous which variables we should
regard as fundamental to be quantized. The reasons we start our
quantum analysis
with this nonlinear $\sigma$-model are two folds. One is that
the original hidden symmetry is manifest in this formulation.
The other is that we can make use of the knowledge of the
quantum theory of nonlinear $\sigma$-models developed in the
literature. Due to this, the quantum analysis of our model is
fairly simplified.

Since the fluctuations to
$x^0$ and $x^3$ directions are ignored, such analysis is not
enough to know the full quantum properties of Einstein gravity.
In particular, we can say only a little about its statistical
aspects. However, we have
at present no consistent way to investigate the full quantum theory
of Einstein gravity because of its nonrenormalizability
and various difficulties.
Our attitude here is a modest one. Although only a part of
the quantum fluctuations can be incorporated, in this simplified
setting we can carry out a consistent quantum analysis of Einstein
gravity and extract some quantum effects on geometry. We believe
that our analysis gives some
insights into quantum aspects of general relativity.
Indeed, it turns out that
we can obtain the forms of the beta functions
to any loop order and
the renormalization effects on the classical solutions.

In order to respect the covariance of the target manifold,
 we rewrite
the action of the nonlinear $\sigma$-model part
by using its metric, $g_{ij} (\phi)$, and coordinates, $\phi^i$,
\eqabegin
  \frac{1}{\hbar} S_{NL} &=& - \frac{1}{2 e_0^2} \int d^2 x  \rho
      g_{ij}(\phi) \del_\mu \phi^i \del^\mu \phi^j
   \comma \label{Snl}
\eqaend
where $e_0^2 = \lp^2/V$ is the coupling of the model.
In string theory, $V$ corresponds to the volume of the compactified
space. On the contrary, in our context $V$ is the "volume"
of the real space-time, $(x^0,x^3)$, and hence $e^2_0$ is an
extremely small number, i.e., the model has a quite small coupling.
The fluctuations depending only on $x^1$ and $x^2$ are constant
modes with respect to the reduced directions, $x^0$ and $x^3$, and
$e_0^2 \propto V^{-1}$ indicates that such fluctuations are
suppressed by the "volume" of the constant direction.
 In the stationary axisymmetric case, which has
the Kerr solution, the time $x^0$ runs from $-\infty$ to
$ +\infty$ and
$V$ tends to infinity. We do not know which value $e_0^2$ takes
in such a case,
but the coupling is still expected to be quite small.

In the coordinates $\phi^{1,2} = \Delta, B $, we have
$g_{ij} = g \delta_{ij}$, $ g = \Delta^{-2} $.
In two dimensions, the curvature tensors are easily calculated
through
\eqabegin
  R &=& - g^{-1} \delta^{ij} \del_i \del_j \ln g \comma \label{R} \\
  R_{ij} &=& \half  R  g_{ij} \comma \label{Ricci} \\
  R_{ijkl} &=&
    \half  R  \left( g_{ik}g_{jl} - g_{il}g_{jk} \right) \period
  \label{Riemann}
\eqaend
{}From the first equation above, we get
\eqabegin
  R &=& -2 \ =\ {\rm const.} \quad \comma
\eqaend
namely, $\coset$ manifold has a negative constant scalar curvature.
Although the above form of $g_{ij}$ appears singular at
$ \Delta = 0 $, this is just a coordinate singularity and we can get
regular metric even there by some appropriate coordinate
transformation.

Now let us consider the renormalization of the effective action
of the model.
As the functional measure, we take
 $\prod_{x, i} \sqrt{{\rm det} g_{ij }} d \phi^i (x)$.
This is invariant under
coordinate transformations of the target manifold, and respects the
covariance.
The only difference
of our model from ordinary nonlinear $\sigma$-models is the
existence of the factor $\rho$ in (\ref{Snl}). This factor
behaves as a coordinate-dependent coupling like dilaton field
in string theory.
In the following, we assume $\rho (x) > 0$.
The reality of the space-time metric
requires just that $\rho (x)$ is real or pure imaginary
(see (\ref{4metric})).
In the case of negative or pure imaginary $\rho$, we have only to
replace $ \rho (x) $ with $ \abs{ \rho (x) } $.
With these in mind, we shall adopt the background field method
and follow \cite{AFM}. Thus our analysis does not depend
on which background we shall take.

First, we expand the action around the background fields,
$\varphi^i$, by normal coordinates,
\eqabegin
   - \frac{1}{\hbar} S_{NL}[\phi]
       & = &  - \frac{1}{\hbar} S_{NL}[\varphi]
      + \int d^2 x
       \ T_0^{-1} g_{ij}(\varphi) \del_\mu \varphi^i D^{\mu}\xi^j
       \nn \\
   && \quad \! + \half \int d^2 x \ T_0^{-1}
  \left[  g_{ij} D_{\mu}\xi^i D^{\mu}\xi^j
          + R_{ik_1 k_2 j} \xi^{k_1} \xi^{k_2}
               \del_\mu \varphi^i \del^\mu \varphi^j
     \right. \label{Snl2} \\
  && \qquad \! \left. + \frac{1}{3} D_{k_1} R_{i k_2 k_3 j}
      \xi^{k_1}\xi^{k_2}
        \xi^{k_3} \del_\mu \varphi^i \del^\mu \varphi^j
     + \frac{4}{3}  R_{i k_1 k_2 k_3}
              \xi^{k_1}\xi^{k_2} D_\mu \xi^{k_3}
        \del^\mu \varphi^i + \cdots \right] \comma  \nn
\eqaend
where $\xi^i$ is the tangent vector to the geodesics around
$\varphi^i$, $T_0(x) = e_0^2/\rho (x)$, and
$D_\mu \xi^i=  \del_{\mu} \xi^i
     + \Gamma^i_{jk} \del_\mu \varphi^k \xi^j$.
$D_k $ is the covariant derivative, e.g.,
$D_k \xi^i = \del_k \xi^i + \Gamma^{i}_{jk} \xi^k$, and
$\Gamma^i_{jk}$ is the Christoffel symbol defined by $g_{ij}$.
Next, we introduce the zweibein, $\hat{h}^p_{\ i}(\varphi, \rho)$,
 with respect to
$\hat{g}_{ij}(\varphi,\rho) \equiv \rho g_{ij}(\varphi)$
  and with
the properties
\eqabegin
  \hat{h}^{p}_{\ i} \ \hat{h}_{p j} = \hat{g}_{ij} &,&
  \hat{h}^j_{\ p} \ \hat{h}_{\ j}^{q}  = \delta_{\ p}^{q} \period
\eqaend
Here the indices for the target manifold, $i,j$, are raised and
lowered by $\hatg_{ij}$ while those for its tangent space,
$p,q$, by $\delta_{pq}$.
(Henceforth we denote the quantities with respect
to $\hatg_{ij}$ by the hat $ \ \hat{} \ $.)
$\hath^p_{\ i}$ is expressed
by the zweibein, $h^p_{\ j}$, with respect to $g_{ij}$ as
$ \hath^p_{\ i} = \sqrt{\rho } \ h^p_{\ i} $.
Then we define $\xi^p = h^p_{\ j} \ \xi^j$ and
$\hatxi^p = \hath^p_{\ i} \xi^i = \sqrt{\rho} \xi^p $. Noting that
$\hath^p_{\ i}$ depends not only on $\varphi$ but also on $\rho$,
we have
\eqabegin
  \hat{h}^p_{\ j} D_\mu \xi^j &=& \calhD_\mu \ \hat{\xi}^p
      \ \equiv  \ \hat{D}_\mu \hat{\xi}^p -
     \frac{1}{2} \del_\mu \ln \rho \cdot
         \hat{\xi}^p
    \comma \label{calD}
\eqaend
where $ \hatD_\mu \hatxi^p
        = \del_\mu \hatxi^p + \hatA^p_{\ q \mu} \hatxi^q$
, $\hatA^p_{\ q \mu} =
          \hat{\omega}^p_{\ q k} \del_\mu \varphi^k$, and
$\hat{\omega}^p_{\ q k} = \hath^p_{\ j}
   \ (\del_k \hath^j_{\ q} + \hat{\Gamma}^j_{kl} \ \hath^l_{\ q})$.
Here we have used $\hat{\Gamma}^i_{jk} = \Gamma^i_{jk}$.
In terms of $\hatxi^p$, the kinetic term has the canonical form,
\eqabegin
  \hatg_{ij} D_\mu  \xi^i D^\mu \xi^j &=&
   \calhD_\mu \hatxi^p \calhD^\mu \hatxi_p
  \ = \ \del_\mu \hatxi^p \del^\mu \hatxi_p + \cdots
  \period
\eqaend

In order to see how other terms are expressed by $\hatxi^p $, we
assign weight $N$ to quantities with the property
$\Phi^{(N)}(\Lambda g_{ij}) =  \Lambda^N \Phi^{(N)}(g_{ij})$
where $\Lambda$ is a constant.
In each term in (\ref{Snl2}), the part without $\xi^i$ and expressed
by the geometrical quantities through $g_{ij}$ has weight $1$
because we are originally expanding
$g_{ij}(\phi) \ \del_\mu \phi \del_\nu \phi$.
Let us denote such a quantity by $\Phi^{(1)}(g_{ij})$.
Since the derivatives of $ \rho $ with respect to $\varphi^k$
vanish, i.e., $\del_k \rho = 0$, it holds that
\eqabegin
   \rho \Phi^{(1)}(g_{ij}) &=&  \rho \Phi^{(1)}(\hatg_{ij}/\rho)
    \ = \  \Phi^{(1)}(\hatg_{ij})
   \ \equiv \ \hat{\Phi}^{(1)}
  \period \label{scale}
\eqaend
For example, $ \rho R_{ijkl}(g_{ij})
 =  R_{ijkl}(\hatg_{ij}) \equiv \hat{R}_{ijkl}$.
{}From Eqs.(\ref{calD}) and (\ref{scale}), we obtain
\eqabegin
  - \frac{1}{\hbar} S_{NL}[\phi]
      & = & - \frac{1}{\hbar} S_{NL}[\varphi]
        + \int d^2 x \  T_0^{-1} g_{ij}(\varphi) \del_\mu \varphi^i
     D^{\mu}\xi^j
       \nn \\
   && \quad  + \frac{1}{2 e_0^2} \int d^2 x
  \left[ \ \calhD_\mu \hatxi^p  \calhD^\mu \hatxi_p
          + \hatR_{ip_1 p_2 j} \hatxi^{p_1} \hatxi^{p_2}
              \del_\mu \varphi^i \del^\mu \varphi^j       \right. \\
  && \qquad  \left. + \frac{1}{3} \hatD_{p_1} \hatR_{i p_2 p_3 j}
      \hatxi^{p_1}\hatxi^{p_2} \hatxi^{p_3}
               \del_\mu \varphi^i \del^\mu \varphi^j
     + \frac{4}{3}  \hatR_{i p_1 p_2 p_3}
          \hatxi^{p_1}\hatxi^{p_2} \calhD_\mu \hatxi^{p_3}
        \del^\mu \varphi^i + \cdots \right] \comma \nn
\eqaend
where $\hatR_{pijk} = \hath^l_{\ p} \hatR_{lijk}$ etc. .
Therefore we find that the changes from the cases without $\rho$
(i.e., ordinary nonlinear $\sigma$-models )
are only (i) the replacement of all the quantities
by those with the hats and (ii) the further replacement
$ \hatD_\mu \hat{\xi}^p \to \calhD_\mu \hat{\xi}^p$.
The term linear in $\xi^i$ contributes to a field redefinition
together with the source term omitted here.
We shall drop this linear term because it is irrelevant to the
following
discussion. Since the transformations,
$\phi^i \to \xi^i \to \xi^p $, are coordinate
transformations on the manifold, the functional measure is
invariant, while
under the last transformation, $ \xi^p \to \hatxi^p$, the measure
is changed into
$\prod_{x, p} \rho^{-1} \ d \hatxi^p (x)$.
When the factor $\rho^{-1}$ is raised into the
action, it is proportional to the delta function. However,
since we shall
adopt dimensional regularization, it plays no role
in the following calculations at least perturbatively
\cite{PZA,Friedan}.

We now proceed to the loop calculations. As long as we are
concerned with divergent parts, we can estimate
the effects due to $\del_\mu \ln \rho$ in $ \calhD_\mu \hatxi^p$
to all orders.
First, let us note that possible counter terms are scalars, and
on dimensional grounds they are of
dimension two and hence including two base-space derivatives.
Second, at $N$-loop order, they have weight $-N + 1 $.
Third, since $\hatR = -2/\rho $ and similar formulae to
(\ref{R})$-$(\ref{Riemann}) are
valid for the quantities with the hats, the covariant derivatives
of the curvatures, $\hatR, \hatR_{ij}$, and $\hatR_{ijkl}$, vanish
and any scalar without the base-space derivative, $\del_\mu$,
is a function
only of $\rho$. Therefore the counter terms at $N$-loop order
has the
factor $T_0^{N-1}$. For example at one and two loop orders,
the possible counter terms including
$\del_\mu \ln \rho$ are proportional to
$\del_\mu \ln \rho \ \del^\mu \ln \rho$ and
$\del_\mu \ln \rho \ \del^\mu \ln \rho \ \hatR$, respectively.
Consequently, we find the counter terms due to
 $\del_\mu \ln \rho$ in $ \calhD_\mu \hatxi^p$ to be of the form
\eqabegin
 \delta S_{NL}^{(\rho)} &=&
     - \frac{1}{4\pi \epsilon}\int d^2 x
     \left( \sum_{N=1}  b_N \ T_0^{N-1} \right)
     \del_\mu \ln \rho \ \del^\mu \ln \rho
 \comma \label{countrho}
\eqaend
where we have adopted the minimal subtraction and the
dimensional regularization, i.e.,
dim.$= 2 \to n$ and $\epsilon = n-2$.
As for the infrared regularization, we have adopted a simple mass
cutoff. Since the renormalization of the model is a problem
concerned with short distances, the scheme of the infrared
regularization
may not be essential.
$b_N$ are numerical coefficients determined by explicit calculations.
It is easy to check $b_1 = 1/2 $. The existence of
$\delta S_{NL}^{(\rho)}$
shows that we have to add an additional bare term,
 $-1/2 \int d^2 x  U_0^{-1} \del_\mu \ln \rho \ \del^\mu \ln \rho$,
in the action, where $U_0 = {\cal O}(e_0^2)$.

As we have already estimated the result from the change
$\hatD_\mu \to \calhD_\mu $, the remaining analysis of the
divergent parts can be performed in a parallel way to ordinary
nonlinear $\sigma$-models.
Thus we immediately get other counter terms up to two loop order
\cite{Friedan,AFM},
\eqabegin
 \delta S_{NL} &=&  \frac{1}{4\pi \epsilon} \int d^2 x \left[
    \hatR_{ij} +  \frac{e_0^2}{4\pi}
             \ \hatR_{iklm} \hatR_j^{\ klm} \right]
        \del_\mu \varphi^i \del^\mu \varphi^j \nn \\
    &=&    \frac{1}{4\pi \epsilon} \int d^2 x
          \left[ - 1 + \frac{T_0 (x)}{2\pi} \right] g_{ij}
                 \del_\mu \varphi^i \del^\mu \varphi^j
   \period \label{twoloop}
\eqaend
Moreover we can determin the form of the remaining counter terms
to all orders. In a similar way to the previous argument, we
find that
any tensor with two lower indices of the target manifold which
is made out of the metric, curvatures and covariant derivatives
are propotional to $g_{ij}$, and that the remaining counter terms
are of the form
\eqabegin
  \delta S_{NL} &=&  \frac{1}{4\pi \epsilon} \int d^2 x
        \left[ \sum_{N=1} \ a_N  T_0^{N-1} \right]
          g_{ij} \del_\mu \varphi^i \del^\mu \varphi^j
   \comma
\eqaend
where $a_N$ are numerical coefficients determined by the explicit
calculations.
(\ref{twoloop}) implies $a_1 = -1$ and $a_2 =1/(2\pi)$.
Note that the sign of
$a_1$ is opposite to usual cases of compact manifolds.

As the counter terms above are functions of $\rho(x)$, the model is
not strictly renormalizable. It is, however, renormalizable in a
more general sense in which the manifold of the classical action
changes due to quantum effects \cite{Friedan}. Indeed, we can derive
beta functions for the couplings,
$T(x;\mu)$ $ \equiv $ $ e^2(x;\mu)/\rho(x) $ and $U(T(x;\mu))$,
as in the usual renormalizable theories \cite{Friedan,AFM}.
They are given by
\eqabegin
  \beta_T \left( T \right) &\equiv& \mu \frac{\del}{\del \mu} T
         \ = \  - \frac{1}{2\pi} \sum_{N=1} N a^{(0)}_N T^{N+1}
   \comma \label{betafn} \\
  \beta_U (T)  &\equiv& \mu \frac{\del}{\del \mu} U
     \ = \   \frac{1}{2\pi} U^2
         \sum_{N=1}\left( N b^{(0)}_{N+1}
             + \del_T \ln U \cdot b^{(0)}_{N} \right) T^N \comma
  \eqaend
where $\mu$ represents the renormalization point, and $a_N^{(0)}$
and $b_N^{(0)}$ are the
lowest part of $a_N$ and $b_N$ in $\epsilon^{-1}$, respectively.
It is easy to integrate the above equations as the beta functions
are just rational functions of $\rho(x)$.

%%%%%%%%%%%%%%%%%%%%%%%%%%%%
%%%%%% section 4 %%%%%%%%%%%
\csection{Quantum Effects on the Kerr Geometry}
In the previous section, we carried out the renormalization of
the model and derived the beta functions of the couplings.
In this section, we investigate
the physical consequences of our quantum analysis.
We are interested in global geometry of space-time, and the
effects of
the higher derivative terms in the effective action are expected to
be small for long distances. Thus we shall focus on the quantum
effects
due to the quadraric derivative terms in the effective action.
Let us here regard $\mu_0$ as representing the energy scale of
the classical theory of the reduced gravity.
Then $ e^2(x;\mu_0) = e^2(\mu_0) $ holds, and the
independent equations of motion
including the quantum effects
become
\eqabegin
  && \del_\mu \del^\mu \rho \ = \ 0 \comma  \label{rho2}\\
  && \Delta \del_\mu \left( T^{-1} \ \del^\mu \Epsilon \right)
     \ = \ T^{-1} \ \del_\mu \Epsilon \del^\mu \Epsilon
    \label{Ernsteq2}\comma \\
  && \del_\zeta \rho \del_\zeta \ln \lambda
         - \half \del_\zeta^2 \rho
     \ = \
    \frac{1}{4} e^2(\mu_0)
      \left( T^{-1} \Delta^{-2}
          \del_\zeta \Epsilon \del_\zeta \bar{\Epsilon}
      + U^{-1} \del_\zeta \ln \rho \del_\zeta \ln \rho \right)
                      \label{emconst2}    \comma \\
  && \Delta^{2} \del_\zeta A  \ = \ i \rho \del_\zeta B
   \label{dual22} \period
\eqaend
Here we adopt a particular choice of the conformal gauge
in two dimensions
represented by $(x^1, x^2)$. As mentioned in Sec.2, we can identify
$\rho(x)$ one of the coordinates since $\rho(x)$ is a free field.
Thus introducing another free field, $z$, conjugate with $\rho$,
we choose the gauge,
\eqabegin
 x^1 = \sigma \rho \comma  &&  x^2 = \sigma z
 \comma \label{rhoz}
\eqaend
 and hence $\zeta = \sigma ( \rho + i z )$, where $\sigma$ is some
constant with dimension of length.
In our context, only the Planck length, $\lp$, is
such a constant made out of the fundamental constants
in the theory. Then we set $\sigma = \lp$.

Now we consider the quantum effects on the Kerr geometry
as an interesting example.
It has been proved that the Kerr geometry is the unique solution
to the stationary axisymmetric Einstein gravity under certain
physical
conditions \cite{Robinson}.
In the following, we set $ t \equiv x^0$ and $\omega \equiv x^3$,
and regard $t$ and
$\omega$ as the time and the azimuthal angle, respectively.
We shall find that the
asymptotically flat region does not undergo any quantum correction,
namely, the asymptotic region is stable. Furthermore,
in a certain approximation at one loop order, it is shown that
the geometry inside
the ergosphere changes considerably no matter how small the
quantum effects may be.

The Kerr solution to Eqs.(\ref{Ernsteq})$-$(\ref{dual2})
are usually expressed by Boyer-Lindquest coordinates,
$r$ (the radial coordinate) and $\theta$ (the polar angle), given by
\eqabegin
    \lp \rho &=&
     x^1  \ = \ \sqrt{r^2 - 2mr + a^2} \ \sin \theta \comma \nn \\
   \lp z &=& x^2 \ = \ \left( r-m \right) \cos \theta
\comma
\eqaend
where $m$ and $a$ turn out to represents the mass and the
angular momentum
per unit mass of the Kerr black hole, respectively.
In these coordinates, the Kerr solution is given by \cite{KSHM}
\eqabegin
  \Epsilon &=&  \Delta + i \ B \comma \nn \\
   && \Delta \ = \ \frac{D - a^2 \sin^2 \theta}{\Sigma}
      \comma \quad B \ = \ \frac{2 m a \cos \theta}{\Sigma} \comma \\
  \lambda^2 &=&
     \frac{D - a^2 \sin^2 \theta}{D + (m^2 - a^2) \sin^2 \theta}
     \comma \quad
   A \ = \ a \frac{2mr \sin^2 \theta}{ D - a^2 \sin^2 \theta}
   \comma
\eqaend
where $D = r^2  - 2 m r + a^2  $
and $\Sigma = r^2 + a^2 \cos^2 \theta$.
Then the line element is written as
\eqabegin
 d s^2 &=& - \left(  1 - \frac{2mr}{\Sigma}\right) d t^2
      + \Sigma \left(  \frac{d r^2}{D} + d \theta^2 \right) \nn \\
    && \qquad - \frac{4mar}{\Sigma} \sin^2\theta \ d \omega \ d t
   + \left(  r^2 + a^2 + \frac{2ma^2r}{\Sigma} \sin^2 \theta \right)
         \sin^2 \theta \ d \omega^2
  \period
\eqaend
The zero of $\Sigma$ and those of $D$
 ( i.e.,  $r = r_\pm  \equiv m \pm \sqrt{m^2 - a^2}$) correspond
to the locations of the curvature singularity and the horizons,
 respectively, while the outer zero of $D - a^2 \sin^2 \theta$
( i.e., $r = r_e \equiv  m + \sqrt{m^2 - a^2 \cos^2 \theta}$ )
represents the outer boundary of the ergosphere.
The asymptotically flat
region is described by $ r $ ( or $ \rho $ ) $\to \infty$.
In this asymptotic region, we have
$(d x^1)^2 + (d x^2)^2 \sim (d r)^2 + r^2 \ (d \theta)^2$, and
$(x^1, x^2)$ represents the flat $2$-plane.
Notice that
$\rho$ tends to vanish as $ r \to r_\pm$ or $ \sin \theta \to 0 $.

Since the beta functions $\beta_T(T)$ and $\beta_U(T)$ are
expanded by the power series
of $T(x;\mu_0)$ $=$ $e^2(\mu_0)/\rho(x) $, the perturbation
is valid except
for the small neighborhoods of order $\lp $ of the axis of the
rotation,
$ \sin \theta = 0 $, and the horizons, $ r = r_\pm$. This means that
the quantum fluctuations become large there.
  It is obvious that in the asymptotic
region the beta function vanishes. Therefore there is no quantum
corrections due to the running couplings in that region and the flat
region remains stable.

In order to further study the physical consequences of our analysis,
we have to solve the equations  (\ref{Ernsteq2})$-$(\ref{dual22}).
The change due to the term
$U^{-1} \del_\zeta \ln \rho \del_\zeta \ln \rho$
in (\ref{emconst2}) can be absorbed into a factor of $\lambda$.
Let us define $f(\rho;\mu)$ and $\lambda_T$ by
$\lambda = f(\rho;\mu) \lambda_T$
and $f(\rho;\mu) \to 1 $ as $\rho \to \infty$. Then taking into
account $\lp \rho = x^1$, we find that $f(\rho;\mu)$ and
$\lambda_T$ are given by
\eqabegin
  && f(\rho;\mu) \ = \ \exp \left( \frac{1}{4} e^2(\mu_0)
   \int_\rho^\infty d \rho'\ \rho'^{-2} U^{-1}(T(\rho';\mu)) \right)
        \comma \\
 && \del_\zeta \rho \del_\zeta \ln \lambda_T
     - \half \del_\zeta^2 \rho
       \ = \
    \frac{1}{4} e^2(\mu_0)
     T^{-1} \Delta^{-2} \del_\zeta \Epsilon \del_\zeta \bar{\Epsilon}
    \period \label{eqlambdaT}
\eqaend
The equation for $\lambda_T$ is of the same form as the
classical equations
for $\lambda$, (\ref{emconst}), up to the replacement $\rho$
with $T^{-1}$.
At a generic order, however, remaining equations are quite
complicated.
Thus, henceforth, we focus on one loop order.
At this order, we have
$T^{-1}(x;\mu) = T^{-1}(x;\mu_0) - (1/2\pi) \ln (\mu/\mu_0)$,
and we can get the solution to (\ref{Ernsteq2}) and
(\ref{eqlambdaT})
from the classical one by the replacements of $\rho$ and
$\lambda $ with
$\rho - e^2(\mu_0)/2\pi \cdot \ln (\mu/\mu_0)  $ and $\lambda_T$.
Unfortunately, by this replacements the last equation
(\ref{dual22}) comes not to meet the integrability condition.
Therefore we shall resort to further approximation.
Here we consider the deviation from the classical solution in the
neighborhood
of $\rho(x) = \rho_0$, and approximate $T^{-1}(x;\mu)$ by
\eqabegin
  e^2(\mu_0) T^{-1}(x;\mu)
    &=& \rho(x) \left\{
  1 - \frac{1}{2\pi} e^2(\mu_0) \rho^{-1}(x) \ln (\mu/\mu_0) \right\}
  \ \sim \  \alpha (\rho_0) \rho(x) \comma
\eqaend
where
$\alpha(\rho_0) =
   1 - e^2(\mu_0)/2\pi \cdot \rho_0^{-1}\ln (\mu/\mu_0) = $const. ,
and it tends to $1$ as $ \mu\to \mu_0$.
This approximation is valid in the region where $\rho(x) >> 1$,
because
$\del_{x^1} \rho^{-1} =
    - \  \lp^{-1} \rho^{-2}$ and $\del_{x^2}\rho^{-1} = 0 $.
In this approximation,
(\ref{Ernsteq2}) is the same as the classical one
and the difference between (\ref{emconst}) and
(\ref{eqlambdaT}) is only the exponents of $\lambda$ and
$\lambda_T$.
Thus all the quantum effects is represented by the change
of $\lambda$, and it is given by
\eqabegin
   \lambda^2 &=& f^2(\rho_0) ( F_1/F_2 )^{\alpha(\rho_0)}
    \comma
\eqaend
where $F_1 = D- a^2 \sin^2 \theta $ and
$F_2 = D + (m^2 - a^2) \sin^2 \theta$.
Therefore we find that in this approximation the geometry becomes
\eqabegin
   d s^2 &=& - \left(  1 - \frac{2mr}{\Sigma}\right) d t^2
      +  \ f^2(\rho_0)
       \left( \frac{F_1}{F_2 }\right)^{\alpha(\rho_0)-1} \Sigma
   \left(  \frac{d r^2}{D} + d \theta^2 \right) \nn \\
    && \qquad - \ \frac{4mar}{\Sigma} \sin^2\theta \ d \omega \  d t
      + \  \left(  r^2 + a^2 +
     \frac{2ma^2 r}{\Sigma} \sin^2 \theta \right)
         \sin^2 \theta \ d \omega^2
 \period
\eqaend
{}From the above expression, we find that the additional zeros and
singularities appear in the metric where $ F_1 $ or
$ F_2 $ vanishes. The conformal properties of
the geometry is very much affected by them.
Moreover we see that these singularities develop curvature
 singularities.
For example, let us consider one of the curvature invariants
defined by
$R_{0303} \equiv
   E^K_{\ 0} E^L_{\ 3} E^M_{\ 0} E^N_{\ 3} R^{(4)}_{KLMN} $.
In our parametrization, it takes the form
$ R_{0303} = \lambda^{-2} F_3(\Delta, B, \rho)$, where $F_3$
is a certain
function of $\Delta$, $B$ and $\rho$. Since, in the case of the
Kerr geometry ( i.e., $\alpha(\rho_0) = 1$ ), it becomes singular
only at $\Sigma = 0 $, $R_{0303}$ comes to
diverge at the zeros of $F_1$ or $F_2$ unless $\alpha(\rho_0) = 1$.
Note that the condition
$\rho (x) >> 1$ holds even there except for the vicinity
of the axis of rotation as long as $m$ and $a$ are large enough
compared with the Planck scale,
and that the outest additional zeros or singularities occur
at the outer boundary of the ergosphere, $r = r_e$.
We need further investigation in order to know whether or not
these singularities are true.
However, our result indicates that
the geometry inside the ergosphere, where unusual phenomena can
take place, is changed considerably
due to the quantum effects. This is the case no matter how
small they may be,
namely, as long as $\alpha(\rho_0) \neq 1$.
%%%%%%%%%%%%%%%%%%%%%%%%%%%%
%\newpage
%%%%%% discussion %%%%%%%%%
\csection{Discussion}
In this article, we studied the quantum theory
 of the Einstein gravity with one time-like and one space-like
Killing vector
formulated as a $\coset$ nonlinear $\sigma$-model. We showed
that the quantum analysis of this model can be carried out
in a parallel way to ordinary nonlinear $\sigma$-models in spite of
the existence of an unusual coupling.
This means that it is possible to investigate consistently the
quantum aspects of Einstein gravity in our limited case.
In consequence, the forms
of the beta functions were determined to all orders up to numerical
coefficients. As an explicit example, we considered the quantum
effects on the Kerr geometry. Then we found that the asymptotically
flat region undergoes no quantum effects and remains stable.
It is also discussed that the inner geometry of the Kerr black hole
is changed considerably.
These
contrast with other quantum approaches to quantum properties of
Einstein gravity,
in which Minkowski space-time becomes unstable,
and/or the solution different much from the classical one
is discarded because of the validity of the perturbations
\cite{Horowitz}-\cite{Natsuume}.

It is obvious that we can deal with
the case with two space-like Killing vectors in the same way,
in which
colliding wave solutions are known. In addition, the extension
to the Einstein-Maxwell system is straightforward, because, when
dimensionally reduced, this system
is also formulated as a nonlinear $\sigma$-model coupled to gravity
as mentioned in Introduction.

Admittedly, our analysis is incomplete to understand the full quantum
properties of Einstein gravity. We can say nothing about the
effects of the truncated degrees of freedom. Even after the
dimensional reduction,
the gravitational part remains to be quantized. We should also study
the effects of the higher derivative terms in the effective action.
In order to investigate the statistical aspects of Einstein gravity,
we have to develop some other approaches.
These are beyond the scope of this article and left as future
problems.
Since we have seen that Einstein's theory is formulated
as a nonlinear $\sigma$-model already in the reduction to three
dimensions,
it may be interesting to consider the application
of three dimensional nonlinear $\sigma$-models.
%%%%%%%%%%%%%%%%%%%%%%%%%%%%
%%%%%% acknowledgment, references %%%%%%%%%%
\csectionast{Acknowledgements}
%\parbigskipn
The author acknowledges helpful discussions
with M. Natsuume, Y. Okawa, T. Yoneya in particular with
I. Ichinose and S. Mizoguchi.
He would like to also thank Y. Kazama for useful comments and
discussions.
The research of the author is supported in part by
JSPS Research Fellowship for Young Scientists (No. 06-4391)
from the Ministry of Education, Science and Culture.

%\newpage
%%%%%%%%%%%%%%%%%%%%%%%%%%%%%%%%%%%%
\def\thebibliography#1{\list {[\arabic{enumi}]}
  {\settowidth\labelwidth{[#1]}\leftmargin\labelwidth
 \advance\leftmargin\labelsep
 \usecounter{enumi}}
 \def\newblock{\hskip .11em plus .33em minus .07em}
 \sloppy\clubpenalty4000\widowpenalty4000
 \sfcode`\.=1000\relax}
\let\endthebibliography=\endlist

\csectionast{ References }

%%%%%%%%%%%%%%%%%%%%%%%%%%%%%%
\end{document}